\documentclass{article}
\usepackage{times,fancyhdr}
\usepackage[dvips]{graphicx}
\usepackage{amsmath,amssymb,bm}
\usepackage{color}
\usepackage{mathrsfs}
\usepackage{setspace}
\usepackage{hyperref}
\usepackage{xcolor,soul}

\makeatletter
\newcommand\be{\@ifstar{\[}{\begin{equation}}}
\newcommand\ee{\@ifstar{\]}{\end{equation}}}
\makeatother

\usepackage{authblk}

\begin{document}
\title{Supermeasured: Violating Bell-Statistical Independence without violating physical statistical independence}
\author[1,*]{Jonte R.\ Hance}
\affil[1]{Quantum Engineering Technology Laboratories, Department of Electrical and Electronic Engineering, University of Bristol, Woodland Road, Bristol, BS8 1US, UK}
\author[2]{Sabine\ Hossenfelder}
\affil[2]{Frankfurt Institute for Advanced Studies, Ruth-Moufang-Str. 1, D-60438 Frankfurt am Main, Germany}
\author[3,$\dagger$]{Tim N.\ Palmer}
\affil[3]{Department of Physics, University of Oxford, UK}
\affil[*]{jonte.hance@bristol.ac.uk}
\affil[$\dagger$]{tim.palmer@physics.ox.ac.uk}
\date{\today}

\maketitle

\begin{abstract}
Bell's theorem is often said to imply that quantum mechanics violates local causality, and that local causality cannot be restored with a hidden-variables theory. This however is only correct if the hidden-variables theory fulfils an assumption called Statistical Independence. Violations of Statistical Independence are commonly interpreted as correlations between the measurement settings and the hidden variables (which determine the measurement outcomes). Such correlations have been discarded as ``fine-tuning'' or a ``conspiracy''. We here point out that the common interpretation is at best physically ambiguous and at worst incorrect. The problem with the common interpretation is that Statistical Independence might be violated because of a non-trivial measure in state space, a possibility we propose to call ``supermeasured''. We use Invariant Set Theory as an example of a supermeasured theory that violates the Statistical Independence assumption in Bell's theorem without requiring correlations between hidden variables and measurement settings (physical statistical independence). 
\end{abstract}

\section{Introduction}
\label{intro}
Bell's theorem \cite{Bell1964OnEPR} has been a milestone in our understanding of quantum mechanics by detailing just what correlations are necessary to reproduce observations. Unfortunately, many physicists have jumped to incorrect conclusions from it. A 2016 survey among professional physicists \cite{Sujeevan2016Survey} found that 34\% believe Bell's theorem shows that ``Hidden variables are impossible,'' that is, they think Bell's theorem rules out theories in which measurement outcomes are determined by variables that are not accounted for in standard quantum mechanics. Similarly, in a survey conducted among professional quantum physicists at a conference in 2012 \cite{Schlosshauer2013Attitudes}, 64\% claimed that Bell's theorem rules out hidden variable theories (and said local realism is untenable). This is of course not so. Bell's theorem merely shows that a hidden variables theory {\emph{which fulfils all the assumptions of the theorem}} is ruled out by observation.

Bell's theorem however contains one questionable assumption: (Bell-)Statistical Independence, sometimes called the ``Free Will'' or ``Free Choice'' assumption (here capitalised and labelled Bell to distinguish it from the intuitive ideas of physical statistical independence, free will and free  choice).
Indeed, one can interpret \emph{all} experiments that have found violations of Bell's inequality as simply demonstrating that if quantum mechanics is underpinned by a local, causal, and deterministic theory, then that underlying theory must violate Bell-Statistical Independence. Clearly the conclusion to draw from this is that we should look for a hidden-variables theory that violates Bell-Statistical Independence, not least to develop a quantum formalism that is compatible with General Relativity. 
Of course this is not historically what has happened. Instead, physicists have collectively discarded the possibility that Bell-Statistical Independence might be violated because they misunderstood what it means. For example, referring to Bell-Statistical Independence as ``free will'' or ``free choice'' seems to have created a strong cognitive bias for accepting the assumption unthinkingly. 

That this ``free will'' nomenclature is highly misleading has already been clarified elsewhere \cite{Hossenfelder2020Rethinking,Hossenfelder2020SuperdeterminismGuide} and we don't want to repeat this entire discussion here (though we will briefly comment on the relation between statistical independence and free will in Section \ref{free}). Our aim here is to investigate the physical interpretation of Bell-Statistical Independence and explain why it is widely misunderstood. 

This misunderstanding is well-illustrated by a quote from a recent paper by Sen \cite{sen2022analysis}:
\begin{quote}
``The [Statistical Independence] assumption states that the hidden variables that determine the measurement outcomes are uncorrelated with the measurement settings.''
\end{quote}
Similar interpretations can be found in \cite{Sen2020Superdet1,Sen2020Superdet2}. This indeed is the standard way of interpreting the mathematical statement of Bell-Statistical Independence. However, we will show below that physically this interpretation is at best ambiguous and at worst wrong. In Section \ref{general}, we give a general argument for this. In Section \ref{CHSH}, we will look at the {\sc CHSH} inequality in particular. In Section \ref{IST}, we will discuss in more detail Invariant Set Theory \cite{Palmer2020Discretization}, which is to our knowledge the only example of a theory that `violates Bell-Statistical Independence without violating physical statistical independence'. Misconceptions regarding free will, fine tuning and conspiracy are discussed in Section \ref{free}.

\section{Understanding Statistical Independence}
\label{general} 

In Bell's theorem, Statistical Independence
is often said to be the assumption that
\be
\label{SI2}
\rho (\lambda | X) = \rho(\lambda)~,
\ee
where $\lambda$ is a set of hidden variables, $X$ are the detector settings, and $\rho$ is a probability distribution of the hidden variables. In Bell's theorem one normally uses two separate detectors and their settings. We will comment on that specifically in Section \ref{CHSH}, but let us first look at the general interpretation. It is possible in principle that $\rho$ depends on further variables, but this won't matter in the following.

That $\rho(\lambda,X)$ is a probability distribution means it is normalised over a space, which we will denote ${\mathscr S}_{\rm math}$ for the mathematical state space: it comprises all mathematically possible states of the hidden-variables theory. By ``mathematically possible'' we literally just mean that we can write them down mathematically. We might however later discard some of the mathematically possible states as not physically possible or meaningful. This isn't so uncommon. For example, some mathematically possible solutions to the Schr\"odinger equation are not normalisable and hence not physically possible. 

The key point we want to make in this section is that any space we integrate over must have a measure, $\mu(\lambda,X)$, and generically this measure is non-trivial, i.e. it isn't just identical to some normalisation constant. A measure roughly speaking quantifies the volume of the space. The probability distribution $\rho$ can only be normalised by help of the appropriate measure:
\begin{equation}
\int_{{\mathscr S}_{\rm math}} \hspace*{-0.5cm} d\lambda dX~ \rho(\lambda, X) \mu(\lambda,X) = 1~.
\end{equation}

Measure theory \cite{MeasureTheory} is not usually discussed in physics textbooks. However, a variety of measures make their appearance in physics nevertheless. The most widely used one is the Lebesgue measure on ${\mathbb R}^n$ and (pseudo-)Riemannian manifolds. On fractals it can be generalised to the Hausdorff measure. In the context of Hamiltonian dynamical systems, a non-trivial measure on state space arises in the theory of symplectic manifolds (leading, for example, to the Gromov non-squeezing theorem). In Section \ref{istmath}, we discuss non-trivial invariant measures associated with chaotic attractors. 

The measure of ${\mathscr S}_{\rm math}$ appears in the calculation of any expectation value and therefore should enter the derivation of Bell's theorem together with the probability distribution $\rho$. Since these two functions always appear together, it is tempting to simply combine them into one $\rho_{\rm Bell} (\lambda,X) := \rho(\lambda,X)\mu(\lambda,X)$, where we use the index ``Bell'' to emphasise that this is the quantity that {really} enters Bell's theorem. 
The assumption of Statistical Independence in Bell's theorem is therefore actually the assumption that
\begin{eqnarray}
\rho_{\rm Bell} (\lambda | X) = \rho_{\rm Bell} (\lambda)~. \label{rhobell}
\end{eqnarray}
We call this Bell-Statistical Independence (with the prefix Bell- and capital letters), and distinguish it from Eq.\ (\ref{SI2}) which we now call physical statistical independence in lower-case letters. (Note, in the most general formulation of the assumptions of Bell's theorem one considers a dependence of $\rho_{\rm Bell}$ on both the measurement settings as well as the preparation procedure. However, the dependence on the preparation procedure is irrelevant to our point, and we have therefore for simplicity omitted it. The reader may consider the preparation details to be part of the hidden variables.)

To avoid confusion with the standard interpretation of superdeterminism, we propose to call a theory which violates Eq.\ (\ref{rhobell}) but does not violate Eq.\ (\ref{SI2}) a ``supermeasured'' theory, with $\mu$ being the supermeasure (note, despite the similarity in terminology, this has nothing to do with measurement).

Since Eq.\ (\ref{rhobell}) is mathematically indistinguishable from Eq.\ (\ref{SI2}) given a suitable redefinition of the probability density, one may wonder why even make the effort of introducing the two distributions $\rho$ and $\rho_{\rm Bell}$? It is important to distinguish them because physically they mean something different. $\rho$ is the distribution of states on ${\mathscr S}_{\rm math}$. It can be affected by factors under the control of the experimenter, such as the preparation of the state. $\rho_{\rm Bell}$, by contrast, is the distribution weighted by the measure $\mu(\lambda, X)$. This measure is not under the control of the experimenter - it's just a property of the laws of physics. As such $\rho_{\rm Bell}$ contains information \emph{both} about the intrinsic properties of the space \emph{and} the distribution over the space. 

The problem with the common interpretation of Bell-Statistical Independence is that typically the measure $\mu$ is not explicitly defined in the assumptions for Bell's theorem. This means that one implicitly assumes that the measure $\mu$ is identical to the uniform measure $\mu_0$ on ${\mathscr S}_{\rm math}$. The consequence is that interpretations of Bell's theorem run afoul of physics whenever one is dealing with a theory in which $\mu(\lambda,X) \ne \mu_0$.

To see why this distinction matters, let us look at a simple idealised example for illustration. The following example is not meant to describe a realistic physical theory. We merely present it to elucidate that it is always possible to replace a correlation on one space with a non-correlated distribution on a subset of the first space -- without changing any of the probabilities. This shows that the common definition of statistical independence is ambiguous for what the physical interpretation is concerned, and so claims of ``fine-tuning'' based on this definition are also ambiguous. We will come to a more physically relevant example later. 

Let ${\mathscr S}_{\rm math}$ be a compact continuous space with uniform measure $\mu \equiv \mu_0 =$~constant, and $\rho$ a probability distribution over it. This probability distribution may violate Eq.\ (\ref{SI2}). Our task here will be to show that we can remove this correlation entirely without changing any probabilities. 

\begin{figure*}[ht]
\centering
\includegraphics[width=\textwidth]{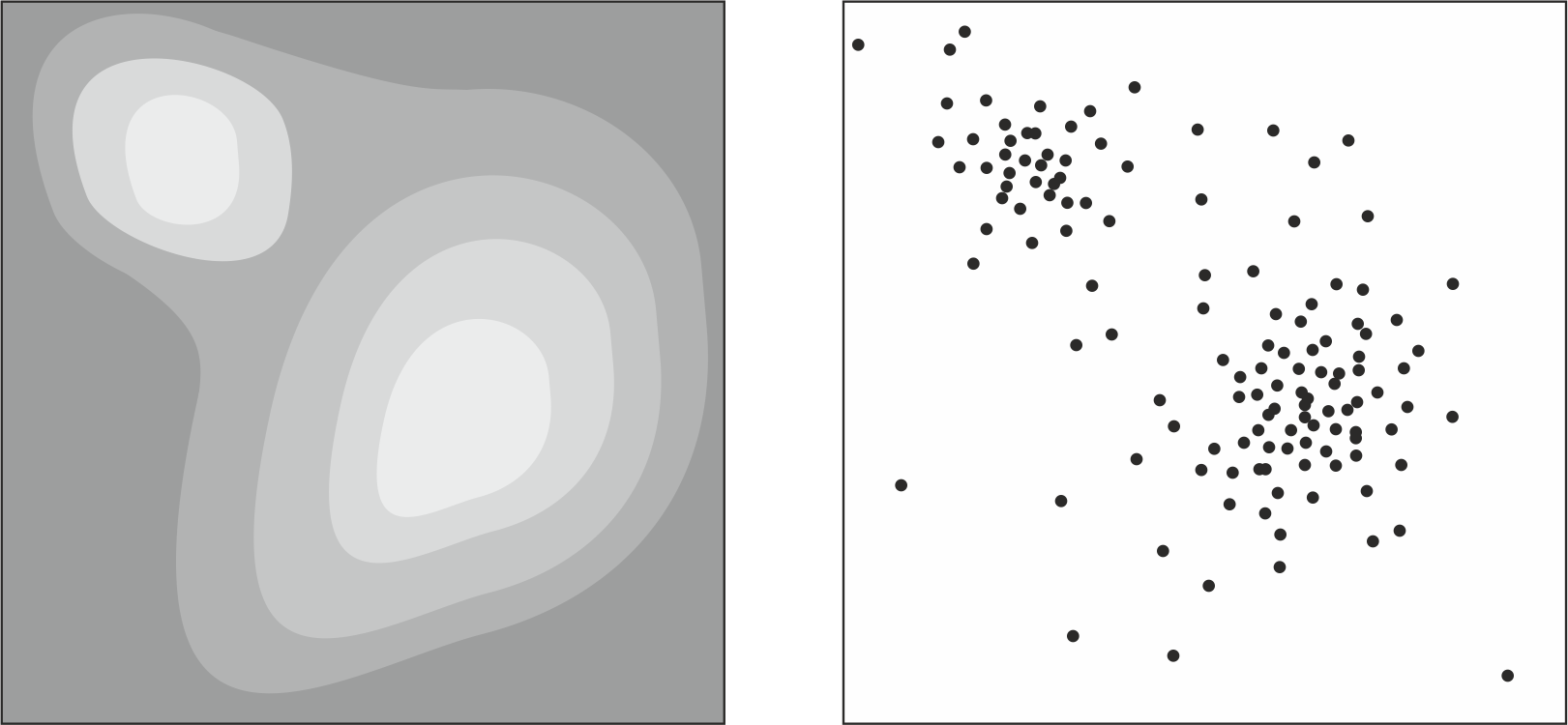}
\caption{Illustration of sampling procedure. Left: The square represents the space ${\mathscr S}_{\rm math}$ over hidden variables and measurement settings, and the shading is the probability distribution $\rho$ over it. The brighter the shading, the higher the probability. Right: We randomly distribute a set of $N$ points using $\rho$. In the limit $N \to \infty$ a uniform distribution $\tilde \rho$ on the points will reproduce the probabilities defined by $\rho$ on ${\mathscr S}_{\rm math}$ with a uniform measure. The set of points defines the new space ${\mathscr S}_{\rm phys}$, also over hidden variables and measurement settings. It has a non-trivial measure $\tilde \mu$ in ${\mathscr S}_{\rm math}$. Any correlations that were present in $\rho$ are thereby moved into the structure of ${\mathscr S}_{\rm phys}$. \protect{\label{fig}}}
\end{figure*}

To see this, we use $\rho$ to randomly choose a set ${\mathscr S}_N = \{ (\lambda_1,X_1),(\lambda_2,X_2)...(\lambda_N,X_N)\}$ of $N$ points in ${\mathscr S}_{\rm math}$. For illustration, see Figure \ref{fig}. From this we define the discrete measure
\begin{eqnarray}
\tilde \mu(\lambda,X) := \sum_{i=1}^N \delta(\lambda - \lambda_i) \delta(X-X_i)~.
\end{eqnarray}
Note that the probability that two points are equal to another is zero.
We then define a uniform probability distribution $\tilde \rho(\lambda,X) \equiv 1/N$ on ${\mathscr S}_{\rm math}$ which is normalized with respect to $\tilde \mu$ and from that $\tilde \rho_{\rm Bell}(\lambda,X):= \tilde \rho(\lambda,X) \tilde \mu(\lambda,X)$.

Let us now take a subset $A$ of ${\mathscr S}_{\rm math}$ with non-zero volume (according to $\mu$), $A \subset {\mathscr S}_{\rm math}$. In the limit $N\to \infty$, we then have for the probability $P(A)$ of finding the system in that subset
\begin{eqnarray}
P(A) = \int_{A} d\lambda dX ~ \tilde \rho_{\rm Bell}(\lambda, X)   = \int_A d \lambda dX ~ \rho_{\rm Bell}(\lambda, X)~,
\end{eqnarray}
for any $A$.
This means that all probabilities calculated from $\rho$ on ${\mathscr S}_{\rm math}$ with uniform measure $\mu \equiv \mu_0$ are by construction identical to those of the uniform distribution $\tilde \rho$ with measure $\tilde \mu$. 

Finally, we define the new, physical state-space ${\mathscr S}_{\rm phys}:= \lim_{N\to \infty} {\mathscr S}_N$. Since $\tilde \mu \equiv 0$ on ${\mathscr S}_{\rm math} \setminus {\mathscr S}_{\rm phys}$, we discard the complement, only keep ${\mathscr S}_{\rm phys}$, and restrict the probability $\tilde \rho$ to ${\mathscr S}_{\rm phys}$.

Once we have done that, the entire information that was previously in $\rho$ has moved into the definition of the physical state-space ${\mathscr S}_{\rm phys}$. 
$\rho_{\rm Bell} = \rho \mu$ and $\tilde \rho_{\rm Bell} = \tilde \rho \tilde \mu$ give exactly the same probabilities. Since we never experimentally measure probability densities but only probabilities, these two theories are physically indistinguishable. Both violate Bell-Statistical Independence as defined in Eq.\ (\ref{rhobell}). But $\tilde \rho(\lambda |X ) = \tilde \rho(\lambda)$, that is, the hidden variables are by construction uncorrelated on the new space. On ${\mathscr S}_{\rm phys}$, the theory would violate Eq.\ (\ref{rhobell}), but not Eq.\ (\ref{SI2}). 

We have here used a uniform distribution on the physically possible states. This corresponds to what is commonly called the principle of indifference \mbox{\cite{keynes1921treatise}}. We just use this as the simplest example of a distribution on state space. Regardless of what the distribution is, however, one cannot be indifferent about the state-space itself because this state space is a property of the laws of physics. It is whatever it is. This can induce violations of Bell-Statistical Independence even if the distribution over the space is uniform.

It is in this sense that the common interpretation of Bell-Statistical Independence is wrong: On a state space with non-trivial measure, the hidden variables may not be correlated with the detector settings and yet the Bell-Statistical Independence assumption will be violated. The common interpretation neglects the possibility that the theory is supermeasured rather than superdeterministic.

Moving the correlation into the definition of the physical state space is a simple way to avoid fine-tuning (the claim that the experimentally confirmed correlations are sensitive to small changes in the distribution (raised e.g. in \cite{Sen2020Superdet2,Wood2015FineTuning})). If the correlations are created by the intrinsic properties of the space itself, rather than the distribution over the space, then small changes just can't happen. This is the idea of Invariant Set Theory ({\sc IST}) \cite{Palmer2020Discretization} which we will discuss in more detail in Section \ref{IST}. Other reasons why the fine-tuning argument can fail were previously discussed in \cite{Hossenfelder2020SuperdeterminismGuide,Wharton2019Reformulations,Donadi2020SuperdetToy}.

Another example of a theory which uses a non-trivial measure is Spekkens toy model \mbox{\cite{Spekkens2007Evidence}}. This model relies on an ``epistemic restriction'' that requires certain combinations of phase-space distributions to have measure zero. The Spekkens model is not supermeasured, however, because this measure does not depend on the measurement setting. For this reason the Spekkens model cannot reproduce Bell-inequality violations, whereas {\sc IST} can.

We also note that one could apply the principle of indifference to say the measure should be uniform (or trivial). This occurs in fields such as statistical mechanics (e.g. with the Gibbs measure \mbox{\cite{georgii2011gibbs}}). However, empirical evidence/observed physics gives us reason to believe the measure is (or at least to consider theories where the measure is) non-trivial, and so the principle of indifference does not apply here. Specifically, we can take clues from the non-commutativity of certain variables in quantum mechanics that we might have gotten the measure of the space wrong, and hence try a different measure to see if it allows us to explain more.

We want to stress, however, that just because a measure \emph{can} remove fine-tuning does not mean it necessarily does. The measure itself may be fine-tuned, in the sense that it requires a large number of details to be specified. Whether that is so must be evaluated for each model on a case by case basis. But since we know already that the measurement settings are sufficient to obtain the correct predictions of quantum mechanics as average values (because that is what we do in quantum mechanics), it is reasonable to think that the measure need not be fine-tuned. We will now show this with a simple example.

\section{Statistical Independence in the CHSH inequality}
\label{CHSH}

Before turning to Invariant Set Theory as a natural example for a non-trivial measure, we will go through a simple example, the {\sc CHSH} inequality \cite{Clauser1969CHSH}.
The {\sc CHSH} setting describes a measurement of two entangled particles with two different detectors, commonly assigned to two observers, Alice ($A$) and Bob ($B$). The inequality states that any locally causal hidden variable theory which respects Bell-Statistical Independence fulfils
\begin{equation}
   \big| E(X_0,Y_0)-E(X_1,Y_0)+E(X_0,Y_1)+E(X_1,Y_1) \big| \leq 2~,
\end{equation}
where $X_{0/1}$ are detector settings on Alice's side and $Y_{0/1}$ are the detector settings on Bob's side. The quantum correlations $E(X_{0/1},Y_{0/1})$ are the expectation values of the results, $A(X_{0/1})$ and $B(Y_{0/1})$, where $A(X_{0/1})$ is Alice's result given Alice's setting $X_{0/1}$, and $B(Y_{0/1})$ is Bob's result given Bob's setting $Y_{0/1}$. 
We will consider the usual case in which there are only two possible measurement outcomes relative to those settings, $A(\cdot),B(\cdot) \in \{-1,+1 \}$.

In experimental tests of the {\sc CHSH} inequality, the correlations for four different combinations of settings are estimated from four separate sub-ensembles of particles. 
Physical statistical independence is then the assumption that 
\be
\label{SI3}
\tilde \rho (\lambda | X_0 Y_0) = \tilde \rho(\lambda | X_0 Y_1) = \tilde \rho(\lambda | X_1 Y_0) = \tilde \rho(\lambda | X_1 Y_1) = \tilde \rho(\lambda)~,
\ee
that is, the distribution of hidden variables does not depend on the (combination of) detector settings. We will assume that our hidden variables theory fulfils this assumption.

We will now explain how the {\sc CHSH} inequality can be violated in a hidden variables model by violating Bell-Statistical Independence but not physical statistical independence. For this, we first assume that the entangled state is represented by a hidden variable, $\lambda$, which has the probability distribution $\tilde \rho(\lambda,XY)$. And next, that the measurement outcome is determined by the hidden variable and the settings: $A(\lambda,X), B(\lambda,Y)$.

We will denote the space of all the hidden variables with $\Lambda$ and divide it up into subsets for each possible combination of detector settings and outcomes $\Lambda_{XY}^{AB}$. That is, the subset $\Lambda^{++}_{00}$ contains all $\lambda$s for setting $X_0Y_0$ that will give the result $A=+1,B=+1$, the subset $\Lambda^{+-}_{00}$ contains all $\lambda$s for setting $X_0Y_0$ that will give the result $A=+1,B=-1$, and so on.

However, we will next assume that the hidden variable cannot occur for two different combinations of measurement settings. If the variable $\lambda$ described the case with setting $X_0Y_0$, then the combination $(\lambda,X_0Y_1)$ is in ${\mathscr S}_{\rm math}$ but not in ${\mathscr S}_{\rm phys}$. This means that 
\begin{eqnarray}
\Lambda = \bigcup_{i,j,k,l} \Lambda^{ij}_{kl}\quad \mbox{for} \quad i,j \in \{+,-\}~\wedge~k,l \in \{0,1\}~,
\end{eqnarray}
but that these spaces are mutually disjoint
\begin{eqnarray}
\Lambda^{ij}_{kl} \cap \Lambda^{ab}_{cd} = \emptyset \quad \mbox{for} \quad ijkl \neq abcd~.
\end{eqnarray}

This does not contradict Eq.\ (\ref{SI3}) because physical statistical independence is a statement about the probability distribution. The probability distributions for the four different combinations of settings can be made similar to arbitrary precision, even though no value of $\lambda$ appears twice. 

Imagine for example that $\lambda$ is a real number from the interval $[0,1] \in {\mathbb{Q}}$. We randomly sample $N$ points from this interval using a uniform distribution and assign each to one of combinations of settings and outcomes, i.e. one of the $\Lambda^{ij}_{kl}$. The so-generated probability distributions will be statistically indistinguishable for $N\to \infty$ even though the probability that $\lambda$ appears for two different settings is zero. Note that this is the case regardless of how many of the $\lambda$'s we assign to each detector provided the cardinality of all subsets is the same. In the limit $N\to \infty$ the distribution for $\lambda$ conditioned on one of the detectors will just be uniform on each subspace $\Lambda^{ij}_{kl}$.

However, and here comes the important point, the measures of the subspaces $\Lambda^{ij}_{kl}$ don't have to be the same. All we need to do now is choose $\tilde \rho$ to be constant, and the measure of the space $\Lambda^{ij}_{kl}$ to be proportional to the quantum mechanical probability $P\Big(A(X_k)B(Y_l)|X_i Y_j\Big)$ (the constant of proportionality will cancel with the normalisation of $\tilde \rho$) for each possible combination of outcomes. Note that, since $\lambda$ together with the detector setting determines the outcome, the outcome isn't an independent variable. As a result, if we want to calculate the expectation value for a certain combination of measurement settings in the hidden variables model, we have
\begin{eqnarray}
E(X_k,Y_l) &=& 
\sum_{AB} \int_{\Lambda^{AB}_{kl}} \hspace*{-0.3cm} d \lambda~ A(\lambda,X_k) B(\lambda,Y_l) \tilde \rho(\lambda) \tilde \mu(\lambda | X_kY_l) \nonumber \\
&=& \sum_{AB} A(X_k) B(Y_l) P\Big(A(X_k)B(Y_l)|X_k Y_l\Big) ~,
\end{eqnarray}
which reproduces the correlations of quantum mechanics.

Again it might seem rather trivial: We have just pushed the quantum mechanical correlation into the definition of the physical state space and then uniformly sampled the hidden variables over this space. This way, physical statistical independence is respected because the correlation comes from the definition of the space rather than from the distribution over it.

The above example can be generalised straight-forwardly to any quantum mechanical measurement, regardless of what variables are measured in what order or how many detectors there are. The above construction will always give the exact same result as quantum mechanics. In particular it will obey the same bounds and violate all other inequalities just the same as quantum mechanics. 

Of course this example is somewhat pointless because we didn't specify the model sufficiently to even know whether it's locally causal. However, any deterministic hidden variables theory that violates local causality can be made locally causal on the expense of violating Bell-Statistical Independence. We have now further shown that -- contrary to what is often stated -- violating Bell-Statistical Independence does not necessarily require correlations between the hidden variables and measurement settings.

This isn't the only way to reproduce quantum mechanics with a locally causal and deterministic model without fine-tuning \cite{Donadi2020SuperdetToy} but it is a nice example to see just why the distribution of the hidden variables is not fine-tuned. It is uniform on the sample-space, and the sample-space just describes what happens in reality. If the detector setting is one thing, it is not also another thing. The correlations come from the sample-space itself. And the information that goes into the construction of the sample-space is just the same as in quantum mechanics. Hence, this model is exactly as fine-tuned or not fine-tuned as quantum mechanics. That is to say, a rational reader who has no quarrels with quantum mechanics should have no quarrels with this model either.

\section{Invariant Set Theory}
\label{IST}

\subsection{The Mathematical Basis of Invariant Set Theory}
\label{istmath}

We next consider supermeasures in the context of Invariant Set Theory ({\sc IST}). We do in the following not directly need the fractal structure of an invariant set. That invariant sets are generically fractals merely serves as a motivation to consider a finite discretisation of Hilbert space in which certain combinations of states do not exist. As we explain below, this discretisation naturally acts as a supermeasure.

Invariant Set Theory ({\sc IST}) \cite{Palmer2020Discretization,Palmer1995Spin,Palmer2009ISP} is a putative theory of quantum physics based on the assumption that the universe is a causal deterministic dynamical system whose state-space is a fractal set, $I_U$. This fractal set corresponds to ${\mathscr S}_{\rm phys}$ of the previous section. It is invariant under the action of dynamical equations: if a point lies on $I_U$, its time evolution always lies on $I_U$; if a point does not lie on $I_U$ its time evolution never will and never has. The nontrivial measure $\tilde \mu$ is the measure of this invariant set. It is sometimes called the invariant measure. Each trajectory of $I_U$ actually comprises a Cantor Set's worth of trajectories. As such $\tilde \mu$ is a Hausdorff measure \cite{Rogers1998Hausdorff} -- a generalisation of Lebesgue measure for spaces with non-integer dimension. 

Notably, if the state space is a fractal, it has gaps. (Indeed one could say it is mostly gaps since the set has measure zero in the continuum embedding space.) In {\sc IST}, the states in the gaps are not ontic, they are counterfactual states that are mathematically possible (and hence lie in ${\mathscr S}_{\rm math}$), but are inconsistent with the assumed laws of physics (and hence do not lie in ${\mathscr S}_{\rm phys}$). With respect to a Euclidean metric on ${\mathscr S}_{\rm math}$, the non-ontic states are arbitrarily close to the ontic states. That is, perturbations (which are tiny with respect to the Euclidean metric) will generically take an ontic state to one that is inconsistent with the assumed laws of physics. This does not make the theory fine-tuned as perturbations which take ontic states off the invariant set are necessarily $p$-large with respect to a $p$-adic norm. Such a norm, and associated metric, is the natural one to use on a fractal geometry. And if {\rm IST} is not fine tuned, it cannot be conspiratorial. 

The fractal structure that underpins $I_U$ is further assumed to be isomorphic to the $p$-adic integers, for some very large $p$. The theory of dynamical systems defined on $p$-adic numbers is an established part of arithmetic dynamics (see e.g. \cite{Woodcock1998padic} and references therein).
Indeed the famous Lorenz model based on three simple ordinary differential equations
\begin{equation}\label{Lorenz}
\begin{split}
&\frac{dX}{dt} = \sigma(Y-X) \\
&\frac{dY}{dt} = X(\rho -Z) -Y\\
&\frac{dZ}{dt} = XY - \beta Z
\end{split}
\end{equation}
provides a motivational example of what we have in mind. No matter where in state space these equations are initialised, the solutions of the Lorenz equations define trajectories which, after an infinite length of time, fall onto the fractal Lorenz attractor. What we are proposing is to base the laws of physics, not on differential equations like Eq.\ (\mbox{\ref{Lorenz}}), but on geometric equations which describe the attractor. With such laws, a point in state space which does not lie on the attractor is not physically consistent with such laws. Such a point is assigned a prior probability of zero. On the other hand, it is impossible to know \textit{a priori} whether a point lies on the attractor: the geometric properties of fractal structures like the Lorenz attractor are formally non-computable \mbox{\cite{blum1998complexity,Dube1993Fractal}}.

{\sc IST} in its current form does not have a dynamical law. However, this is commonplace in the quantum foundational literature as in many cases one only cares about transition amplitudes between initial and final times. Those amplitudes are in addition often between spin states, so that one does not need to consider a space-time evolution. A typical example of this is Spekkens' Toy Model \mbox{\cite{Spekkens2007Evidence}}, which does not have a dynamical evolution equation but despite that has proved to be useful. Like in Spekkens' Toy Model, we study here what insights we can extract directly from the structure of state space.

As a consequence of this fractal structure, an ensemble-based probabilistic state of the system in {\sc IST} can be expanded in the basis of detector eigenstates $|A_j \rangle$ in the form:
\be\label{Eq:Expansion}
|\psi\rangle = a_1 |A_1\rangle + a_2|A_2\rangle \ldots + a_J |A_J\rangle~,
\ee
where the coefficients as usual square to 1, ie $\sum_j a_j a^*_j =1$ and the star denotes complex conjugation.
The crucial difference to standard quantum theory is that in {\sc IST} the complex amplitudes $a_j$ belong to a subset of the complex numbers $a_j \in \mathbb C_p, \mathbb C_p \subset \mathbb C$. The elements of $\mathbb C_p$ obey rationality restrictions on the coefficients (and so do not form a field). Specifically, if we write $a_j$ in polar form $a_j=R_j e^{i \phi_j}$ then the fractal structure of $I_U$ demands
\begin{eqnarray}
\label{Eq:finite}
R^2_j=m_j/p; \ \ \ 
\phi_j = 2\pi n_j/p~,
\end{eqnarray}
where $m_j,n_j,p \in {\mathbb N}_0$, $m_j,n_j < p$.

This discretisation is effectively a nontrivial measure, allowing the violation of Bell-Statistical Independence -- it gives measure zero to any states whose coefficients in Eq.\ (\ref{Eq:Expansion}) do not obey the conditions in Eq.\ (\mbox{\ref{Eq:finite}}). This means distributions over the set of allowed states will violate the assumption of Bell-Statistical Independence, even if the distributions themselves contain no information about the detector setting.

In the limit $p \rightarrow \infty$, the set of such ``rational'' Hilbert states is dense in the projection of the quantum mechanical Hilbert-space. From this point of view, for large enough $p$, {\sc IST} can be made as experimentally indistinguishable from quantum theory as one likes. This makes it difficult to devise experimental tests for this idea. Such tests must ultimately be based on the fact that, at the end of the day, $p$ is some finite number \cite{Hance2021ExpIST}. 

However, no matter how large is $p$, the state-space of this theory will continue to have gaps. That is to say, the limit $p \rightarrow \infty$ is singular: quantum mechanics does not correspond to {\sc IST} in the large $p$ limit. Importantly, no matter how large is $p$, ensembles of trajectories described by Hilbert States where the complex amplitudes do not belong to $\mathbb C_p$, do not lie on $I_U$. As such, these trajectories have the measure $\tilde \mu=0$ in the continuous embedding space. 

With this {\sc IST} explains why it is impossible to simultaneously measure conjugated variables in quantum mechanics with certainty. In quantum theory, this is a consequence of having non-commuting operators acting on a Hilbert-space, but is otherwise unexplained. In {\sc IST} this arises in a deterministic framework because of the geometric structure of the invariant set and associated fractal measure. The incomplete algebraic structure of ${\mathbb C}_p$ reflects the ``gappy'' geometric structure of $I_U$. For example, superpositions of two states which are in ${\mathbb C}_p$ are generically not also in ${\mathbb C}_p$. 

Of course, thinking about rational numbers and constraints among which states can mutually exist does not explain all results of quantum mechanics -- that would require, amongst other things, a dynamical law. It does however provide a mathematical basis for the impossibility of measuring certain combinations of variables at the same time.

\subsection{The CHSH Inequality in Invariant Set Theory}
\label{CHSHIST}

It was previously demonstrated in \mbox{\cite{Palmer2020Discretization}}, that {\sc IST} correctly reproduces the observed violations of Bell's inequality and the results of sequential Stern-Gerlach experiments. We will here use the {\sc CHSH} inequality to explain how the measure of state space is relevant to obtain the correct probabilities.

In {\sc IST}, if we keep the setting of the first detector fixed (say, at $X_0$), then the settings $Y_0$ and $Y_1$ of the second detector cannot both be on the set if $Y_0 \neq Y_1$. If $Y_0$ was on the set together with $X_0$, then $Y_1$ won't be on it, or the other way round. One of them is not physically possible. This property is a consequence of the rationality requirement on the amplitudes and phases and a number-theoretic result known as Niven's theorem
 \cite{Niven1956irrational,Jahnel2010Sines}:
\begin{quote}
{\bf Niven's Theorem:}
 \emph{Let $\phi/2\pi \in \mathbb{Q}$. Then $\cos \phi \notin \mathbb{Q}$ except when $\cos \phi =0, \pm \frac{1}{2}, \pm 1$.}
\end{quote}
If the first combination of settings, ($X_0Y_0$) for given $\lambda$, fulfils the rationality condition, then the second one, $(X_0Y_1)$ for the same $\lambda$, can't fulfil it; it is associated with a state in ${\mathscr S}_{\rm math}$ where $\tilde \mu=0$. 

This means that 
\begin{eqnarray}
\tilde \rho_{\rm Bell} (\lambda | X_0 Y_0) \ne 0 \implies \tilde \rho_{\rm Bell}(\lambda | X_0 Y_1) = \tilde \rho_{\rm Bell}(\lambda | X_1 Y_0)=0 \\
\tilde \rho_{\rm Bell} (\lambda | X_1 Y_0) \ne 0 \implies \tilde \rho_{\rm Bell}(\lambda | X_0 Y_0) = \tilde \rho_{\rm Bell}(\lambda | X_1 Y_1)=0
\end{eqnarray}
etc. However, this does not imply
\begin{eqnarray}
\tilde \rho(\lambda | X_0 Y_0) \ne 0 \implies \tilde \rho(\lambda | X_0 Y_1) = \tilde \rho(\lambda | X_1 Y_0)=0\\
\tilde \rho(\lambda | X_1 Y_0) \ne 0 \implies \tilde \rho(\lambda | X_0 Y_0) = \tilde \rho(\lambda | X_1 Y_1)=0
\end{eqnarray}
etc.

Now because the statistics of trajectories in {\sc IST} can be represented by complex Hilbert vectors over $\mathbb C_p$ for large $p$, the measures of the spaces $\Lambda^{ij}_{kl}$ introduced in Section \ref{CHSH} are proportional to the quantum mechanical probabilities $P\big(A(X_k)B(Y_l)|X_k Y_l\big)$. Hence {\sc IST} reproduces the correlations of quantum mechanics.

\subsection{Free Will}\label{free}
As mentioned at the beginning, Bell-Statistical Independence is sometimes called the Free Will and/or Free Choice assumption. This refers to the notion that if, for example, Alice actually chose $X_0$ and Bob $Y_0$, then, keeping $\lambda$ and Alice’s choice fixed, Bob could have chosen $Y_1$ instead -- he had the freedom to have done otherwise. But as we have also discussed, such a counterfactual state is incompatible with {\sc IST}: If Bob actually chose $Y_0$, then the state corresponding to the triple $(\lambda, X_0, Y_1)$ is not in the physical state-space; it corresponds to a point in ${\mathscr S}_{\rm math}$ where $\tilde \mu=0$. 

But does this constraint actually have anything to do with free choice? In a deterministic framework, like {\sc IST}, free will can at best be interpreted as the absence of constraints that could prevent an agent from doing as they please. But there is nothing in {\sc IST} that would prevent an agent from doing as he or she pleases any more than this is always the case in any deterministic theory -- the laws of nature always constrain what we can do. And just as it is possible to violate Bell-Statistical Independence without violating physical statistical independence, it is possible to violate Free Choice (the assumption in Bell's Theorem) without violating free choice -- remember that after all Free Choice is just a fancy name for Bell-Statistical Independence. That is, in {\sc IST} Bob can freely choose \emph{among the physically possible options} in the sense that there is no constraint on them. 


While the difference between the correlations being on state space or in the probability distributions makes little difference to our observable world, it has relevance to the difference between Free Will as an assumption in Bell's Theorem, and ``free will'' as debated in metaphysics. Were the ``free will'' debate based on what we observe or experience, we would be tempted to say that moving correlations to a supermeasure makes no difference; however, most of the free will debate is metaphysical -- it has nothing to do with what we observe or experience. It concerns the question of what we even mean by being ``free'' from something. And for that part of the debate, it matters whether an event or process is even physically possible (in the physical state space) or merely mathematically possible. 

For instance, no one seems to ever be worried that we are not free to move around with complex-valued velocities (or momenta). Mathematically, this is totally possible. It just does not happen in reality -- observables are Hermitian operators. However, no one has ever argued that this restricts our ``free will''. Why not? Because that velocities are real-valued is just a property of how the universe is. Obtaining a complex-valued velocity is not a physically possible change, so we do not think it restricts our ``free will''. (It is probably something most people do not think about in the first place.) It is for this reason that the distinction between mathematical and physical possibilities matters, even though this does not affect what actually happens experimentally.

In some parts of the literature, authors have tried to distinguish two types of theories which violate Bell-SI. Those which are superdetermined, and those which are retrocausal. The most naive form of this (e.g. \mbox{\cite{sen2022analysis}}) seems to ignore the prior existence of the measurement settings, and confuses a correlation with a causation. More generally, we are not aware of an unambiguous definition of the term ``retrocausal'' and therefore do not want to use it.

In the supermeasured models that we consider, the distribution of hidden variables is correlated with the detector settings at the time of measurement. The settings do not cause the distribution. We prefer to use  find Adlam's terms -- that superdeterministic/supermeasured theories apply an ``atemporal'' or ``all-at-once'' constraint -- more apt and more useful \mbox{\cite{adlam2022two}}.

\section{Conclusion}

While Bell's theorem is often said to imply that local causality (which is violated by standard quantum mechanics) cannot be restored with a deterministic hidden variables theory, this is only correct if the hidden-variables theory respects Bell-Statistical Independence. Violations of Bell-Statistical Independence are commonly interpreted as implying a correlation between the measurement settings and the hidden variables which determine the measurement outcomes. However, as we have shown here, one can violate the (Bell-)Statistical Independence assumption in Bell's theorem without any correlations between the measurement outcomes and the hidden variables. The violations of Bell-Statistical Independence can instead come about by the geometry of the underlying state space. We have argued that this is a simple way to see that violating Bell-Statistical Independence does not require fine tuning.

\bigskip

\textit{Acknowledgements---}
We thank Sophie Inman and John Rarity for useful discussions. TP is funded by a Royal Society Research Professorship. SH acknowledges support by the Deutsche Forschungsgemeinschaft (DFG, German Research Foundation) under grant number HO 2601/8-1. JRH is supported by the University of York's EPSRC DTP grant EP/R513386/1, and the EPSRC Quantum Communications Hub (funded by the EPSRC grants EP/M013472/1 and EP/T001011/1).

\bibliographystyle{unsrturl}
\bibliography{ref.bib}

\end{document}